\documentclass[tightenlines,floatfix,nofootinbib,showpacs,aps,epsf,amsfonts,amssymb,amsbsy]{revtex4}

\usepackage{graphics,bm}
\usepackage{epsfig}
\usepackage{graphicx}

\newcommand{\beq}{\begin{equation}}
\newcommand{\eeq}{\end{equation}}
\newcommand{\beqa}{\begin{eqnarray}}
\newcommand{\eeqa}{\end{eqnarray}}
\newcommand{\egamma}{\gamma}
\newcommand{\neweqline}{\nonumber \\ &&}

\newcommand{\lagrd}{{\cal L}}
\newcommand{\half}{{1\over2}}
\newcommand{\psibar}{{\bar{\psi}}}
\newcommand{\sumint}{\left.\sum\!\!\!\!\!\!\!\!\!\int\right.}

\def\sumint{\hbox{$\sum$}\!\!\!\!\!\!\int}
\def\square{\vcenter{\vbox{\hrule height.5pt
          \hbox{\vrule width.5pt height5pt
          \kern5pt\vrule width.4pt}\hrule height.5pt}}}

\begin{document}

\author{Jens O. Andersen} 
\email{andersen@tf.phys.no}
\author{Rashid Khan}
\email{rashid.khan@ntnu.no}
\author{Lars T. Kyllingstad}
\email{Lars.kyllingstad@ntnu.no}

\affiliation{Department of Physics, 
Norwegian Institute of Science and Technology, N-7491 Trondheim, Norway}
\title{The chiral phase transition and the role of vacuum fluctuations}

\date{\today}

\begin{abstract}
We apply optimized perturbation theory to the quark-meson model
at finite temperature $T$ and quark chemical potential $\mu$.
The effective potential is calculated to one loop both in the
chiral limit and at the physical point and used 
to study the chiral dynamics of two-flavor QCD.
The critical temperature and the order of the phase transition depends
heavily on whether or not 
one includes the bosonic and fermionic vacuum fluctuations 
in the effective potential. A full one-loop
calculation in the chiral limit predicts  a first-order transition 
for all values of $\mu$. At the physical point, 
one finds a crossover in the whole $\mu-T$ plane.

\end{abstract}

\maketitle


%
\section{Introduction}
\label{intro}
    If we consider QCD in the chiral limit, 
    the QCD Lagrangian
    has an extended symmetry.
    For $N_f$ quark flavors, the Lagrangian is
    invariant under $SU(N_f)_L \times SU(N_f)_R$ transformations.
    In the vacuum, this symmetry is spontaneously broken down 
    to $SU(N_f)_V$ by the formation of a chiral condensate.
    According to Goldstone's theorem, this gives rise to
    $N_f^2-1$ massless modes, the Nambu-Goldstone bosons.


    In the real world, quarks are not massless.  If we add a mass term
    for the quarks
    to the QCD Lagrangian, chiral symmetry is explicitly broken,
    which means that the Goldstone bosons become massive as well.
    Nevertheless, the $u$ and $d$ quarks are light enough that we
    have an approximate chiral symmetry in Nature, and for
    simplicity, we therefore often consider them to be massless.
    QCD with two flavors of massless quarks is a commonly used
    approximation, the validity of which seems to be supported by
    the observed lightness of the pions.
    However, there are more than two quark flavors in the real world.
    In particular, the strange quark mass is on the order of 100 MeV,
    and so we expect it to have an impact on physics at this
    energy scale.

    If one heats strongly interacting matter, it is expected that
    there will be a transition to a phase where chiral symmetry is 
    restored and the quarks within the hadrons are liberated.
    The location and nature of the transition from the
    phase with broken chiral symmetry to the chirally symmetric phase
    is a topic of active research.
    The order of the phase transition depends on
    the number of quark flavors and their masses.
    For example, for two quark flavors and vanishing baryon number chemical
    potential $\mu=0$, we expect, a second-order
    phase transition in the chiral limit if the $U(1)_A$ is explicitly broken,
    otherwise it is driven first order.
    In the case of finite quark masses it is a crossover.
    For three massless quarks, $N_f=3$, 
    the transition is always first order.
    These results follow from 
    universality-class arguments and the (non)existence of stable fixed 
    points~\cite{pisarski,stephanov}.
    At nonzero baryon chemical
    potential and in particular at $T=0$, the order of the chiral transition
    is not obvious from universality 
    arguments~\cite{qcdmu}. However, most model calculations predict a 
    first-order
    transition at $T=0$~\cite{stephanov,buballa}. 
    If this is the case, there is a  
    line of first order chiral transitions in the $\mu$-$T$ plane
    that ends in a second-order transition.

    Currently, the only way to obtain quantitative information about
    the chiral phase transition from first principles is to use
    lattice Monte Carlo simulations. 
    Results from lattice QCD indicate that
    the transition takes place at $T_c \sim 150$~MeV depending on the number
    and masses of the quarks~\cite{hotqcd,wuppertal}.

    Unfortunately, lattice QCD is severely limited by the
    sign problem.  When the baryon number chemical potential
    is nonzero, the fermion determinant, which is used as
    the statistical weight in the Monte Carlo importance sampling,
    becomes complex.
    There are several methods to sidestep the sign problem for small
    $\mu$.
    Examples include reweighting techniques, Taylor expansions
    in $\mu$ around $\mu=0$,
    and the use of imaginary chemical potentials.
    For a review, see Ref.\ \cite{Philipsen:2010gj}.
    However, these methods only work for $\mu \lesssim T$,
    and a general solution to the sign problem for all $\mu$
    has yet to be found.

    A complementary approach to the investigation of the chiral phase
    transition, which works even at nonzero baryon number chemical
    potential, is offered by effective models---simplified
    models that share QCD's chiral symmetry breaking pattern.
    Examples include 
    linear 
    sigma models~\cite{Chiku:1998kd,lenaghan,Chiku:2000eu,Herpay:2006vc,Farias:2008fs},
    NJL-type 
    models~\cite{Scavenius:2000qd,kneur,costa2,kahara,nickel}, 
    Polyakov-NJL models~\cite{costa}
    the quark-meson (QM)
    effective 
    model~\footnote{The QM model is also known as the
    linear sigma model coupled to quarks (LSMq). }
~\cite{Scavenius:2000qd,jakobi,mocsy,bj,Schaefer:2006ds,Kovacs:2006ym,jocke,Bowman:2008kc,Skokov:2010sf,Jakovac:2010uy,koch}, (see also~Ref.~\cite{Fraga:2009pi}),
    and the Polyakov-QM model~\cite{kahara,fuku,Schaefer:2007pw,mats,mao,Marko:2010cd,gupti,Herbst:2010rf,nonlocal}.
    In this paper we will employ the QM 
    model.

    By definition, when using effective models we neglect
    certain aspects of the full theory.  In $\chi$PT and
    sigma models, for instance,
    we only have mesonic degrees of freedom.  For this reason,
    there is no way to include a nonzero baryon number chemical potential
    in these models.

    In NJL-type models we have quark degrees of freedom, however, the
    color symmetry is global, which means that there are no gluon
    degrees of freedom.  One-gluon exchange has been replaced by
    local four-fermion interactions between the quarks.
    The NJL model is not confining and the quarks are not asymptotically 
    free~\cite{buballa}. 
    Unlike the scalar models above, it allows the inclusion
    of separate chemical potentials for different quark flavors
    (and colors), and through the use of NJL models a rich structure
    of color superconducting phases at $\mu \neq 0$ and low
    temperature has emerged~\cite{alford}. 

    While the effective models themselves are already simplifications
    of a more complicated theory, one often makes additional simplifications
    within the models.  For qualitative considerations, for instance,
    one may decide to omit contributions 
    that do not seem
    immediately important for the problem at hand.
    As an  example, in the QM model, chiral symmetry breaking takes place in
    the meson sector, and as a consequence, the vacuum contribution
    to the free energy from the fermions
    is sometimes omitted. 
    In the
    NJL model, on the other hand, this term is responsible for the
    chiral symmetry breaking, and so it cannot be neglected.

    When making such simplifications, however, it is important to
    know exactly what one is discarding, in order not to
    ``throw the baby out with the bath water''.  To wit, for the example
    just given, it was recently shown that neglecting the fermion
    vacuum contribution to the QM free energy changes
    the order of the phase transition \cite{Skokov:2010sf}.
    In the present paper we will study the role of the vacuum fluctuations
    in detail and generalize to $\mu \neq 0$.

    The quark-meson effective model contains, as the name
    implies, both quark and meson degrees of freedom.  The
    mesons are described as in the linear sigma model, by an
    $O(N)$-symmetric scalar field theory with a quartic
    self-interaction.  The meson sector
    couples to the quark sector through Yukawa-type couplings.

    We shall investigate the thermodynamics of the QM model using
    optimized perturbation theory (OPT), which was introduced
    in 1998 by Chiku and Hatsuda \cite{Chiku:1998kd,Chiku:2000eu}.
    In this framework, a local quadratic term proportional
    to an arbitrary parameter $\chi$ is added to and subtracted from
    the 
    Lagrangian~\cite{Chiku:1998kd,Chiku:2000eu,yuk,steve,duncan1,duncan2,kleinert,russ,janke,karsch,Andersen:2000yj}
In the perturbative expansion, the added mass term is 
    included in the free part of the Lagrangian
    and the subtracted term is treated as an interaction term.
    This constitutes a reorganisation of the perturbative series in
    which higher-order terms of na\"ive perturbation theory are
    resummed to all orders.
    
    We have already mentioned that the pions are the
    Nambu--Goldstone bosons associated with the spontaneous
    breaking of chiral symmetry.
    As such, the Nambu--Goldstone theorem plays a significant
    role in the study of the chiral phase transition.
    Importantly,
    OPT ensures that this theorem is satisified order-by-order 
    of the expansion \cite{Chiku:2000eu}.

    The paper is organized as follows:
    In Sec.~\ref{sec:model}, we present the quark-meson effective
    model and the OPT framework;
    in Sec.~\ref{sec:thermodynamics}, we calculate the free
    energy to one loop;
    in Sec.~\ref{sec:results}, we present our results
    and in Sec.~\ref{sec:conclusion}, we make a few concluding
    remarks.
    Finally, expressions for various sum-integrals are
    provided in an appendix.

\section{Quark-meson model} \label{sec:model}
    We consider the quark-meson effective model with $N$ real scalar fields
    ($\phi_1, \phi_2, \ldots, \phi_N$) and $N_f$ massless quark flavors
    ($\psi$) with $N_c$ colors.
    The bare Euclidean Lagrangian for the model can be written in the form
    \beq
        \lagrd_B = \lagrd + \Delta\lagrd,
    \eeq
    where $\lagrd$ is the renormalized Lagrangian and $\Delta\lagrd$
    contains the counterterms.  For clarity, we further write $\lagrd$ as
    \beq \label{eq:ren_lagrd}
        \lagrd =
            \lagrd_\mathrm{LSM}
          + \lagrd_\mathrm{q}
          + \lagrd_\mathrm{Yukawa}.
    \eeq
    The first term in this expression, $\lagrd_\mathrm{LSM}$, is the
    Lagrangian of the linear sigma model:
    \beqa 
        \lagrd_\mathrm{LSM} &=&
            \half (\partial_\mu \phi_i)(\partial_\mu \phi_i)
          + \half m_0^2 \phi_i \phi_i
          + \frac{\lambda}{4!} (\phi_i \phi_i)^2
          - h \phi_1\;.
    \eeqa
    In the case of vanishing $h$, this Lagrangian has an $O(N)$
    symmetry which, assuming $m_0^2 < 0$, is spontaneously broken
    down to $O(N-1)$.  When $h \neq 0$ the symmetry is
    explicitly broken, mimicking the explicit breaking of chiral
    symmetry in the QCD Lagrangian by nonzero quark masses.

    Next, we have $\lagrd_\mathrm{q}$, describing the massless quarks:
    \beq \label{eq:quark_lagrangian}
        \lagrd_\mathrm{q} =
         \bar{\psi}
 (\egamma_\mu \partial_\mu - \mu \egamma_4) \psi\;.
    \eeq
    Here, $\mu$ is the quark number chemical potential, which
    we assume to be equal for all quark flavors, and which is
    related to the baryon number chemical potential by $\mu = \mu_B/3$.
    $\lagrd_\mathrm{q}$ has a $U(1)_B$ symmetry related to
    the conservation of baryon number, as well as an
    $SU(N_f)_L \times SU(N_f)_R$ chiral symmetry.

    In Eq.~(\ref{eq:quark_lagrangian}) and throughout the remainder
    of the paper, we take $\gamma_\mu$ to denote Euclidean
    gamma matrices.
    These are related to the familiar gamma matrices 
in Minkowski space
by
$        \egamma_i \equiv i\gamma_M^i$ and
$        \egamma_4 \equiv \gamma_M^0$.
    They have the following properties:
 $       \{\egamma_\mu, \egamma_\nu\} = 2 \delta_{\mu\nu} \mathbf{1}$
and 
$        {\rm tr}[\egamma_\mu \egamma_\nu] = 4 \delta_{\mu\nu}
$.

    $\lagrd_\mathrm{Yukawa}$ describes the Yukawa couplings between
    the scalars and the quarks, and we write it compactly in the form
    \beq
          \lagrd_\mathrm{Yukawa} =
              g \psibar \Gamma_i \psi \phi_i.
    \eeq
    Here, we have defined
    \beq
        \Gamma \equiv(1, -i\egamma_5 {\bf \tau}) ,
    \eeq
    where ${\bf\tau} = (\tau_1, \tau_2, \tau_3)$ are the
    Pauli matrices.  This only makes sense when
    $N=4$, which is the value we will use in the end.
    The fifth Euclidean gamma matrix is defined as
$        \egamma_5 \equiv \egamma_1 \egamma_2 \egamma_3 \egamma_4 = -\gamma_M^5.
$
    Finally, the counterterm Lagrangian $\Delta\lagrd$ is written as
    \beqa
        \Delta\lagrd &=& 
            \half A (\partial_\mu \phi_i)(\partial_\mu \phi_i)
            + \half B m_0^2 \phi_i \phi_i
            + \frac{C \lambda}{4!} (\phi_i \phi_i)^2
            - D m_0^4 
            + F\psibar\egamma_\mu \partial_\mu  \psi
 \label{eq:ct_lagrd}
            + G g \psibar \Gamma_i \psi \phi_i\;,
    \eeqa
    where $A$, $B$, $C$, $D$, $F$, and $G$ are the renormalisation
    constants for the scalar field, the scalar mass,
    the four-scalar coupling, the vacuum energy, the fermion field,
    and the Yukawa coupling, respectively. In the present calculation, we need
to know $B$, $C$, and $D$ to leading order in the couplings.

\subsection{Broken symmetry}
    The group $O(4)$ is homomorphic to $SU(2)\times SU(2)$.
    The consequence of this is that if we take $N=4$ and $N_f=2$,
    an $O(4)$ rotation among the scalar fields can always be
    countered by an $SU(2)\times SU(2)$ transformation on
    the fermions, keeping the entire Lagrangian invariant.

    Furthermore, $O(3)$, the symmetry group of $\lagrd_\mathrm{LSM}$
    in the broken phase, is homomorphic to $SU(2)$, which is
    the symmetry group of $\lagrd_\mathrm{q}$ after chiral symmetry
    has been broken by, say, the addition of a nonzero mass
    term.  Indeed, we shall see that spontaneous symmetry breaking
    in the mesonic sector in fact induces a dynamic mass term
    for the fermions.

    In the phase with broken chiral symmetry, the scalar field $\phi$
    takes on a nonzero vacuum expectation value, and provided $h > 0$, the
    vacuum state will always point in the $\phi_1$ direction.
    When $h = 0$, $O(N)$ symmetry allows us to choose any direction
    for the vacuum state, and for simplicity we choose the same direction
    in this case.
    We therefore write
    \beq
        \phi = (\sigma + v, {\bf\pi})\;,
    \eeq
    where $\sigma$ and ${\bf\pi}$ are fluctuations representing
    the sigma meson and the pions, while $v$ is the vacuum expectation
    value of $\phi_1$.

\subsection{Optimized perturbation theory}
    In optimized perturbation theory (OPT), we add and subtract quadratic
    terms in the Lagrangian:
    \beq
        \lagrd \to
            \lagrd
          + \half \chi \sigma^2 + \half \chi \pi_i \pi_i
          - \half \chi \sigma^2 - \half \chi \pi_i \pi_i\;.
    \eeq
    The first two terms are treated as part of the free Lagrangian 
    and make this explicit
    by absorbing them into a new mass parameter $m$, defined by
    \beq
        m^2 \equiv m_0^2 + \chi\;.
    \eeq
    The subtracted terms are treated as two-particle interactions.
    Optimized perturbation theory is then defined by the counting rule
    that 
    an insertion of $\chi$ counts as a factor of $\lambda$ or $g^2$
    in loop diagrams.    

    Note that we must take care to adjust the counterterms as well.
    For instance,
    \beq
        \Delta\lagrd_0 =
            \half B (m^2 - \chi) v^2
            + \frac{C \lambda}{4!} v^4
            - D (m^2 - \chi)^2\;.
    \eeq

    The optimization parameter $\chi$ is in principle completely
    arbitrary, and if we could carry out calculations to all
    orders, the results would be independent of $\chi$.
    In order to complete a calculation, we therefore
    need a prescription for $\chi$.
    There are infinitely many possible choices, but we
    limit ourselves to mentioning the following 
    two prescriptions:
    \begin{itemize}
        \item The \emph{principle of minimal sensitivity} (PMS) says that
            one should try to minimize the dependence of the result on the
            parameter $\chi$.  Thus, a PMS condition on some observable
            calculated to $\ell$ loops, $\mathcal O_\ell$, would then be 
    expressed as
            \begin{equation}
                \frac{d \mathcal O_\ell}{d m} = 0\;.
            \end{equation}
        \item A \emph{fastest apparent convergence} (FAC) condition means
            that the perturbative corrections in $\mathcal O_\ell$ should be
            as small as possible.
            Sometimes it is even possible to enforce the strict condition
            that
            \begin{equation}
                \mathcal O_\ell - \mathcal O_{\ell-n} = 0
            \end{equation}
            for some $n$ satisfying $1 \leq n \leq \ell$.
    \end{itemize}

\section{Free energy} \label{sec:thermodynamics}
    We now calculate the free energy $\mathcal F$ to one loop.
    In the following, we employ the notation
    \beqa
        \sumint_P &\equiv&
            \left(\frac{e^\gamma \Lambda^2}{4\pi}\right)^\epsilon
            T \sum_{P_0=2\pi nT} \int_{p}{d^{d}p\over(2\pi)^d}\;,\\
\nonumber
        \sumint_{\{P\}} &\equiv&
            \left(\frac{e^\gamma \Lambda^2}{4\pi}\right)^\epsilon
            T \sum_{P_0=(2n+1)\pi T+i\mu} \int_{p}{d^dp\over(2\pi)^d}\;,
\\ &&
    \eeqa
    for the dimensionally regularized sum-integral, where $d=3-2\epsilon$
    is the dimension of space and $\Lambda$ is an arbitrary momentum
    scale. The factor $e^\gamma/4\pi$ is chosen so that, after subtraction
    of the poles in $\epsilon$, $\Lambda$ coincides with the
    $\overline{\mathrm{MS}}$ renormalization scale.

    To make the various contributions to the prefactors of the diagrams as
    explicit as possible, we take the number of scalar fields to be $N$
    (i.e.\ $(N-1)$ pion fields) and the number of quarks to be $N_f$.

    At tree-level, the free energy is given by
    \beqa
\label{treee}
        \mathcal F_0 &=& 
            \half m^2 v^2 + \frac{\lambda}{4!} v^4 - h v\;.
    \eeqa
    To one-loop order, we obtain
    \beq
        {\cal F}_1 =
            {\cal F}_{\rm 1a}
            + {\cal F}_{\rm 1b}
            + {\cal F}_{\rm 1c}
            - \half \chi v^2
            + \Delta {\cal F}_1\;,
    \eeq
    where
    \beqa
        {\cal F}_{\rm 1a} &=&
            \frac{1}{2} \sumint_P \log(P^2 + m_\sigma^2)\;,
            \label{eq:f_1a_diagram}
        \\
        \mathcal F_{1b} &=&
             \frac{1}{2} (N-1) \sumint_P \log(P^2 + m_\pi^2)\;,
            \\ \nonumber
        \mathcal F_{1c} &=&
             -  N_cN_f \sumint_{\{P\}}
                {\rm tr}\log(i\egamma_\mu P_\mu + m_q - \mu \egamma_4)\;,
\\ &&
            \label{eq:f_1c_diagram}
    \eeqa
    and $\Delta \mathcal F_1$ is the one-loop counterterm:
    \beq
\label{1div1}
        \Delta \mathcal F_1 =
            \half B_1 m^2 v^2
            + \frac{C_1 \lambda}{4!} v^4
            - D_1 m^4\;.
    \eeq
The masses are
\beqa
\label{eq:m_sigma}
m_{\sigma}^2&=&m^2+{\lambda\over2}v^2\;,\\
\label{eq:m_pi}
m_{\pi}^2&=&m^2+{\lambda\over6}v^2\;,\\
\label{eq:m_q}
m_{q}^2&=&g^2v^2\;.
\eeqa

    The relevant sum-integrals are given in Appendix \ref{sec:sumint}.
    Expanding the temperature-independent terms to order $\epsilon^0$, we find
    \beqa
        \mathcal F_{1a} &=&
            - \frac{1}{4 (4 \pi)^2}
            \left(\frac{\Lambda}{m_\sigma}\right)^{2\epsilon}
            \bigg\{
                \left[
                    {1\over\epsilon}
                    + \frac{3}{2}
                \right] m_\sigma^4
                + 2 J_0(\beta m_\sigma) T^4
            \bigg\},
\label{1aexp}
        \\ 
        \mathcal F_{1b} &=&
            - \frac{(N-1)}{4 (4 \pi)^2}
            \left(\frac{\Lambda}{m_\pi}\right)^{2\epsilon}
            \bigg\{
                \left[
                    {1\over\epsilon}
                    + \frac{3}{2}
                \right] m_\pi^4
                + 2 J_0(\beta m_\pi) T^4
            \bigg\},
        \\ 
        \mathcal F_{1c} &=&
            \frac{N_cN_f}{(4 \pi)^2}
            \left(\frac{\Lambda}{m_q}\right)^{2\epsilon}
            \bigg\{
                \left[
                    {1\over\epsilon}
                    + \frac{3}{2}
                \right] m_q^4
                - [K^+_0(\beta m_q, \beta\mu) + K^-_0(\beta m_q, \beta\mu)] T^4
            \bigg\}\;, 
\label{1cexp}
    \eeqa
where $J_0(\beta m)$ and $K_0^{\pm}(\beta m_q,\beta\mu)$ are defined in Appendix
A.
    The divergent terms in $\mathcal F_{1a-1c}$ are thus
    \beqa
        \mathcal F_{1a-1c}^\mathrm{div} &=&
            - \frac{1}{4 (4 \pi)^2 \epsilon}
            \left[ m_\sigma^4 + (N-1) m_\pi^4 
\hspace{2cm}
-4N_cN_f m_q^4 \right]
\;.
\label{1a1c}
    \eeqa
    In the minimal subtraction scheme the counterterm $\Delta \mathcal F_1$
    cancels exactly the divergences in the free energy~(\ref{1a1c}),
    which yields the one-loop counterterms 
    \beqa \label{eq:1loop_counterterms}
        B_1 &=& \frac{N+2}{6 (4\pi)^2} \frac{\lambda}{\epsilon},
        \\
        C_1 &=& \frac{1}{(4\pi)^2 \epsilon} \left[
            \frac{1}{6} (N+8)\lambda
            - 24 N_cN_f \frac{g^4}{\lambda}
            \right],
\\ 
        D_1 &=& -\frac{N}{4 (4\pi)^2} \frac{1}{\epsilon},
\label{eq:1loop_counterterms2}
    \eeqa
    The final result for the renormalized one-loop free energy is 
obtained by adding Eqs.~(\ref{1aexp})--~(\ref{1cexp}) and Eq.~(\ref{1div1})
using Eq.~(\ref{eq:1loop_counterterms})-(\ref{eq:1loop_counterterms2})
    \beqa
        \mathcal F_1 &=&
            -\half \chi v^2 - \frac{1}{4 (4\pi)^2} \bigg\{
                \left[L_\sigma+ \frac{3}{2} 
 \right] m_\sigma^4
                + \left[L_\pi + \frac{3}{2}  
\right] (N-1) m_\pi^4
                - \left[L_q +\frac{3}{2}   \right] 4 N_cN_f m_q^4
        \neweqline 
                + 2 \left[
                    J_0(\beta m_\sigma)
                  + (N-1) J_0(\beta m_\pi)
                  + 2N_cN_f K^+_0 (\beta m_q,\beta\mu)
                  + 2 N_cN_f K^-_0 (\beta m_q,\beta\mu)
                \right] T^4
            \bigg\}\;,
\label{f11}
    \eeqa
    where we have defined
    \beq
        L_a\equiv \log\frac{\Lambda^2}{m_a^2},
        \qquad a = \sigma,\pi,q\;.
    \eeq
The full one-loop free energy ${\cal F}_{0+1}$ is then given by the sum of
Eqs.~(\ref{treee}) and~(\ref{f11}). Using the renormalization group
equation for the running coupling $\lambda$, one can show that 
${\cal F}_{0+1}$ is renormalization group invariant.

\section{Results and discussion} \label{sec:results}
    In this section we discuss the procedure for performing
    numerical calculations using OPT, and we present our results.





    At one-loop order, the PMS condition on the free energy,
    \beq
        \frac{d \mathcal F_{0+1}}{d\chi} = 0\;,
    \eeq
simply has no solution.

    Instead, we consider the gap equation for $v$, obtained by differentiating
    the free energy:
    \beq \label{eq:oneloop_gap}
        \frac{d \mathcal F_{0+1}}{d v} = v (m_\pi^2 + \Pi_1) - h = 0.
    \eeq
    Here, $\Pi_1$ is the one-loop pion self-energy at zero external momentum
    \beq
        \Pi_1 = {1\over v} \frac{d \mathcal F_1}{d v}.
    \eeq
    A FAC condition on the self-energy is then $\Pi_1 = 0$, 
    or~\footnote{In some cases $\Pi_1=0$ has no solution. We then minimized
    $|\Pi_1|$ instead in accordance with the FAC condition.}
    \beqa 
\chi &=&        
            \frac{\lambda}{2} R(m_\sigma, T)
          + \frac{N-1}{3} \; \frac{\lambda}{2} R(m_\pi, T)
          + 4 N_cN_f g^2 S(m_q, T, \mu),
\label{eq:chi_full}
    \eeqa
    where
    \beqa\nonumber
        R(m, T) &\equiv&
          - \frac{m^2}{(4 \pi)^2} \left[L_m+1\right]
          + \frac{T^2}{(4 \pi)^2} J_1 (\beta m),
\\ &&
        \\ \nonumber
        S(m, T, \mu) &\equiv&
            \frac{m^2}{(4 \pi)^2} \left[ L_m+1\right]
          + \frac{T^2}{2 (4 \pi)^2}
            [K_1^+(\beta m, \beta \mu) + K_1^-(\beta m, \beta \mu)].
    \eeqa
    This condition on $\chi$ has a straightforward physical
    interpretation---it means that Goldstone's theorem is
    fulfilled even at tree level:
    \beq \label{eq:oneloop_goldstone}
        m_\pi^2 = \frac{h}{v}.
    \eeq
    This also means that we can write $m_\sigma^2 = h/v + \lambda v^2/3$,
    and thus all $\chi$-dependence is removed from the r.h.s.\ of
    Eq.\ \ref{eq:chi_full} in the broken phase.

    In order to perform numerical calculations, we need to find
    numerical values for the model parameters $m_0$, $\lambda$,
    $g$, and $\Lambda$ in terms of the vacuum ($T=\mu=0$) values
    of $m_\sigma$, $m_\pi$, $m_q$, and $v$.  
    Matching at tree level 
    yields~\footnote{At tree level, it is the combination
    $m^2=m_0^2+\chi$ that is determined, cf. 
    Eqs.~(\ref{eq:m_sigma})--(\ref{eq:m_pi}).}
    \beqa
        m_0^2+\chi&=& - \half (m_\sigma^2 - 3 m_\pi^2)\;, \label{eq:vacuum_m0} \\
        \lambda &=& \frac{3}{v^2} (m_\sigma^2 - m_\pi^2)\;, \\
        g       &=& \frac{m_q}{v}\;, \\
        h       &=& m_\pi^2 v.
\label{hhh}
    \eeqa
At tree level, $\Lambda$ is undetermined.
We choose the renormalization scale to be
    \beqa
        \Lambda^2 &=&{m_{\sigma}^2\over e}\;.
        \label{eq:vacuum_Lambda}
    \eeqa
Finally, we set $N=4$, $N_f=2$, and $N_c=3$.


\subsection{Contributions to $\mathcal F_{0+1}$}
    At the one-loop level we distinguish between the following
    contributions to the free energy:
    \begin{enumerate}
        \item The tree-level contribution from the bosons,
            \beq
                \mathcal F_b^\mathrm{tree}
                \equiv
                \mathcal F_0
                =
                \half m^2 v^2 + \frac{\lambda}{4!} v^4 - h v.
            \eeq
        \item The vacuum contribution from the bosons,
            \beqa
                \mathcal F_b^\mathrm{vac}
                &\equiv& 
                  - {1\over 4 (4 \pi)^2} \bigg\{
                        \left[ L_\sigma +\frac{3}{2} 
\right]
                        m_\sigma^4
                      + \left[L_{\pi}+ \frac{3}{2} \right]
                        (N - 1)
                        m_\pi^4
                    \bigg\}\;.
            \eeqa
        \item The thermal contribution from the bosons,
            \beqa\nonumber
                \mathcal F_b^T
                &\equiv&
                  - \frac{T^4}{2 (4 \pi)^2} \left[
                        J_0(\beta m_\sigma)
                      + (N - 1) J_0(\beta m_\pi)
                    \right].
            \eeqa
        \item The vacuum contribution from the fermions,
            \beq
                \mathcal F_f^\mathrm{vac}
                \equiv
                    \frac{N_cN_f m_q^4}{(4 \pi)^2}
                    \left[ L_q+ \frac{3}{2} 
 \right].
            \eeq
        \item The thermal contribution from the fermions,
            \beqa
                \mathcal F_f^T
                &\equiv&
                  - \frac{N_cN_f T^4}{(4 \pi)^2} \left[
                        K_0^+(\beta m_q, \beta \mu)
                      + K_0^-(\beta m_q, \beta \mu)
                    \right]\;.
            \eeqa
    \end{enumerate}
    (There is also the OPT term, $-\half\chi v^2$, which we omit
    for the purpose of the following discussion.  It is taken into
    account in all numerical calculations.)
    We will now investigate the effect of these various contributions
    on the structure of the phase diagram.

\subsection{Chiral limit}
    We start by considering the phase diagram in the chiral limit, $h=0$,
    where Lagrangian has an exact chiral symmetry and the pions are
    massless Nambu--Goldstone bosons.
    We then fix our model parameters using the vacuum values
    \beqa
        m_\sigma &=& 600\textrm{ MeV}\;,\\
        m_\pi    &=& 0\;, \\\
        m_q      &=& 300\textrm{ MeV}\;,\\
        v        &=& f_\pi = 93\textrm{ MeV}\;,
    \eeqa
    and using Eqs.\ (\ref{eq:vacuum_m0})--(\ref{eq:vacuum_Lambda}), we
    find that
    \beqa
        m_0^2+\chi &=& -(424.26\textrm{ MeV})^2\;, \\
        \lambda &=& 124.87\;, \\
        g &=& 3.2258\;, \\
        h &=& 0\;,\\
        \Lambda &=& 363.92\textrm{ MeV}\;.
    \eeqa

\begin{figure}[htb]
\includegraphics[width=7cm]{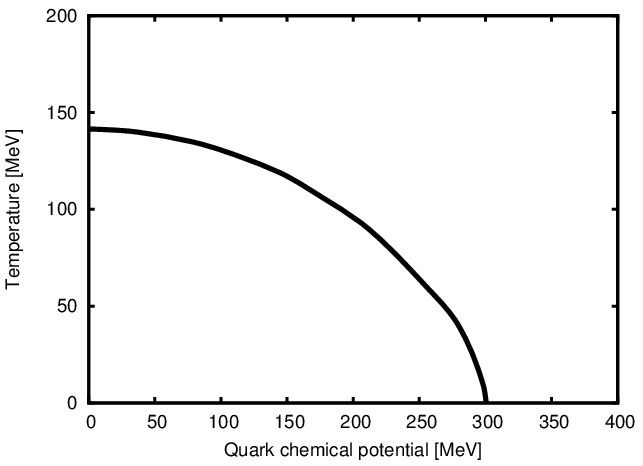}
\includegraphics[width=7cm]{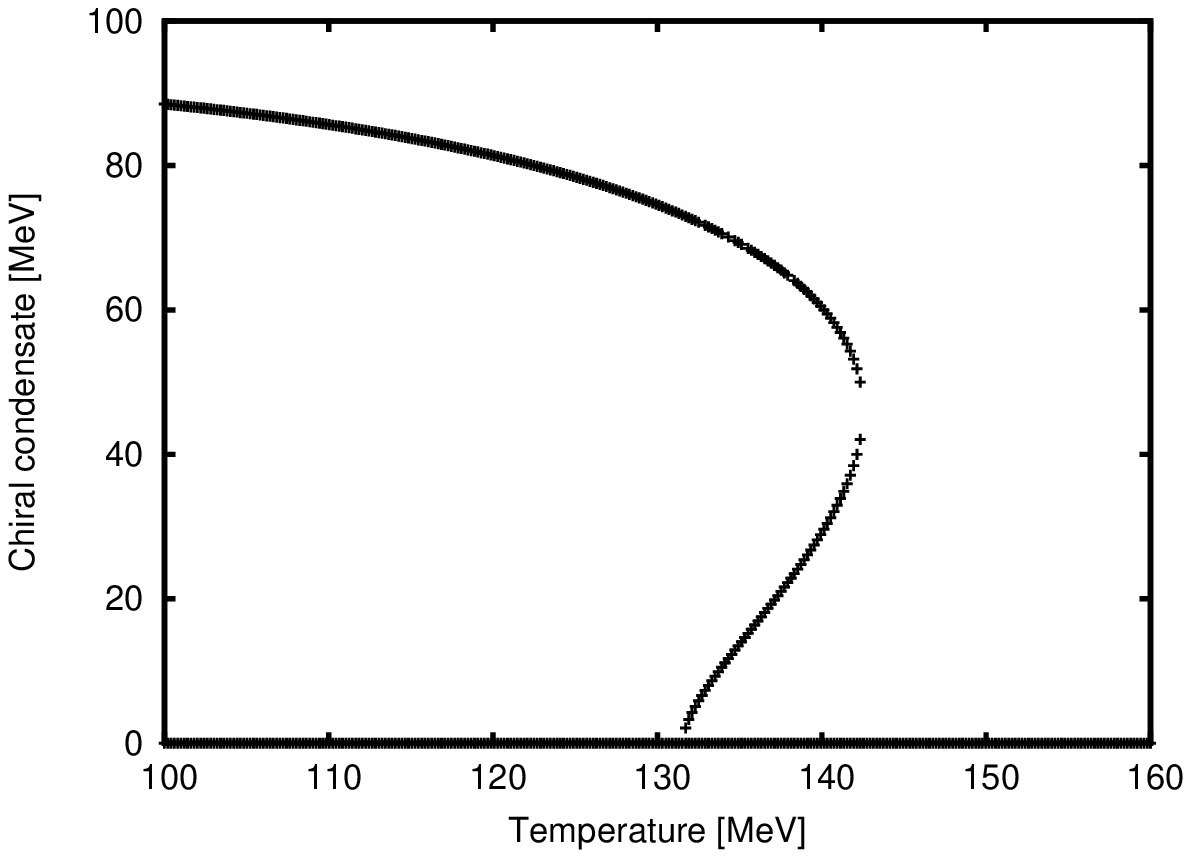}
        \caption{
            At one loop, including only the thermal contribution of the
            fermions and the tree-level contribution of the bosons:
            \emph{(left panel)} 
the phase diagram in the $\mu-T$ plane, where the
            solid line represents a first-order phase transition,
            and
            \emph{(right panel)} the solutions of the gap equation plotted as a
            function of $T$ at $\mu = 0$.
            \label{fig:1-loop_chiral-limit_no-vacuum_no-bosons}
        }
    \end{figure}

    Starting with the simplest case, we may say that we are only
    interested in thermal effects, and thus ignore the vacuum
    contributions $\mathcal F_b^\mathrm{vac}$ and
    $\mathcal F_f^\mathrm{vac}$.  Furthermore, it is common
    to use a mean-field approximation to the meson sector,
    in which we ignore the thermal fluctuations, $\mathcal F_b^T$,
    as well.
    Hence, all that remains of the free energy is the
    tree-level boson contribution and the thermal fluctuations
    of the quarks:
    \beq
        \mathcal F =
            \mathcal F_b^\mathrm{tree}
          + \mathcal F_f^T.
    \eeq
    The phase diagram one obtains using this approximation is shown
    in Fig.\ \ref{fig:1-loop_chiral-limit_no-vacuum_no-bosons} 
({\it left panel}).
    We see that the phase transition is first order for all values
    of $\mu$, a fact which is explicitly demonstrated for $\mu=0$
    in the right panel.  The critical
    temperature is $T_c = 140$ MeV, which is reasonably close to
    the critical temperature of QCD.
    The critical chemical potential for $T=0$ is $\mu_c=296$ MeV.

\begin{figure}[htb]
        \includegraphics[width=7cm]{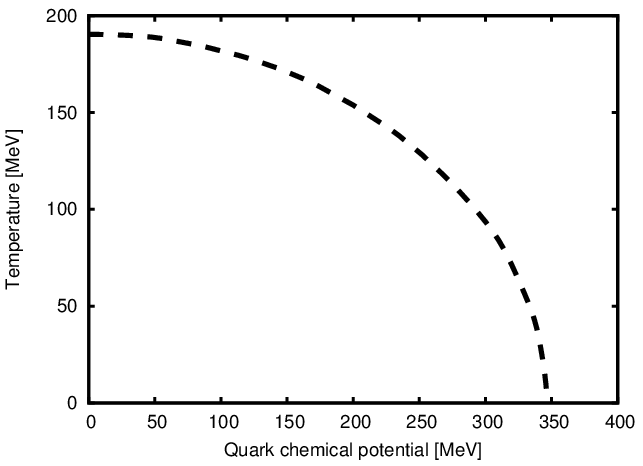}
        \includegraphics[width=7cm]{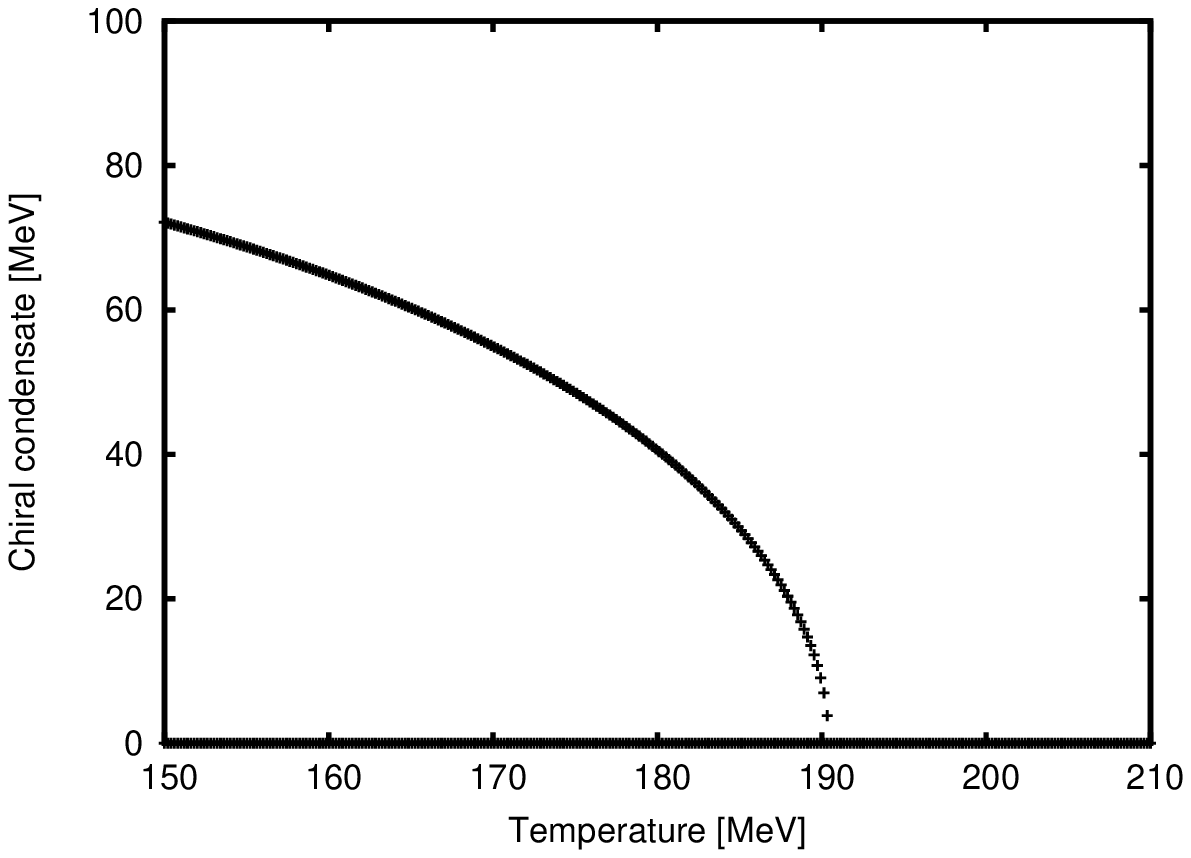}
        \caption{
            At one loop, including both the vacuum contribution and the
            thermal contribution of the fermions, but only the tree-level
            contribution of the bosons:
            \emph{(left panel)} the phase diagram in the $\mu-T$ plane, where 
            the
            dashed line represents a second-order phase transition,
            and
            \emph{(right panel)} the solutions of the gap equation plotted as a
            function of $T$ at $\mu = 0$.
            \label{fig:1-loop_chiral-limit_no-bosons}
        }
    \end{figure}

    Recently, Skokov {\it et al}~\cite{Skokov:2010sf} showed 
analytically for $\mu = 0$ 
that if one
    includes the fermion vacuum term, i.e.
    \beq
        \mathcal F =
            \mathcal F_b^\mathrm{tree}
          + \mathcal F_f^\mathrm{vac}
          + \mathcal F_f^T,
    \eeq
    the order of the phase transition changes---it becomes
    a second-order phase transition. For $T=0$ and finite $\mu$, this
    can be shown analytically as well.
    This is displayed in the right panel of
    Fig.\ \ref{fig:1-loop_chiral-limit_no-bosons}.
    Furthermore, as the left panel shows, the phase transition
    is of second order for all values of $\mu$. 
    The critical temperature at $\mu=0$ is $T_c = 190$ MeV, a significant
    increase as compared to the case where the fermion vacuum fluctuations were 
    ignored. 
    Similarly, there is a large increase in
    the critical chemical potential for $T=0$ to $\mu_c=345$ MeV.

\begin{figure}[htb]
        \includegraphics[width=7cm]{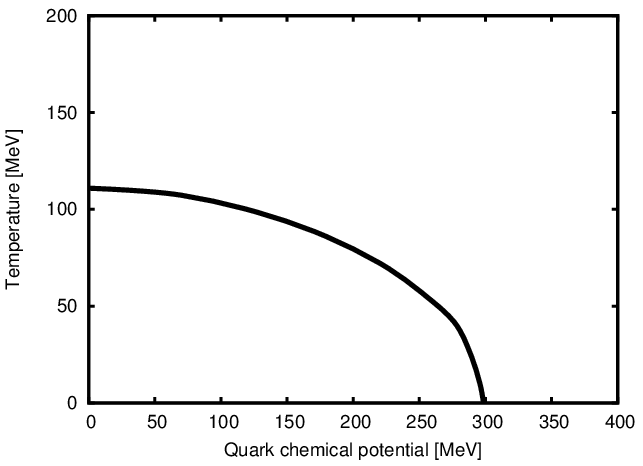}
        \includegraphics[width=7cm]{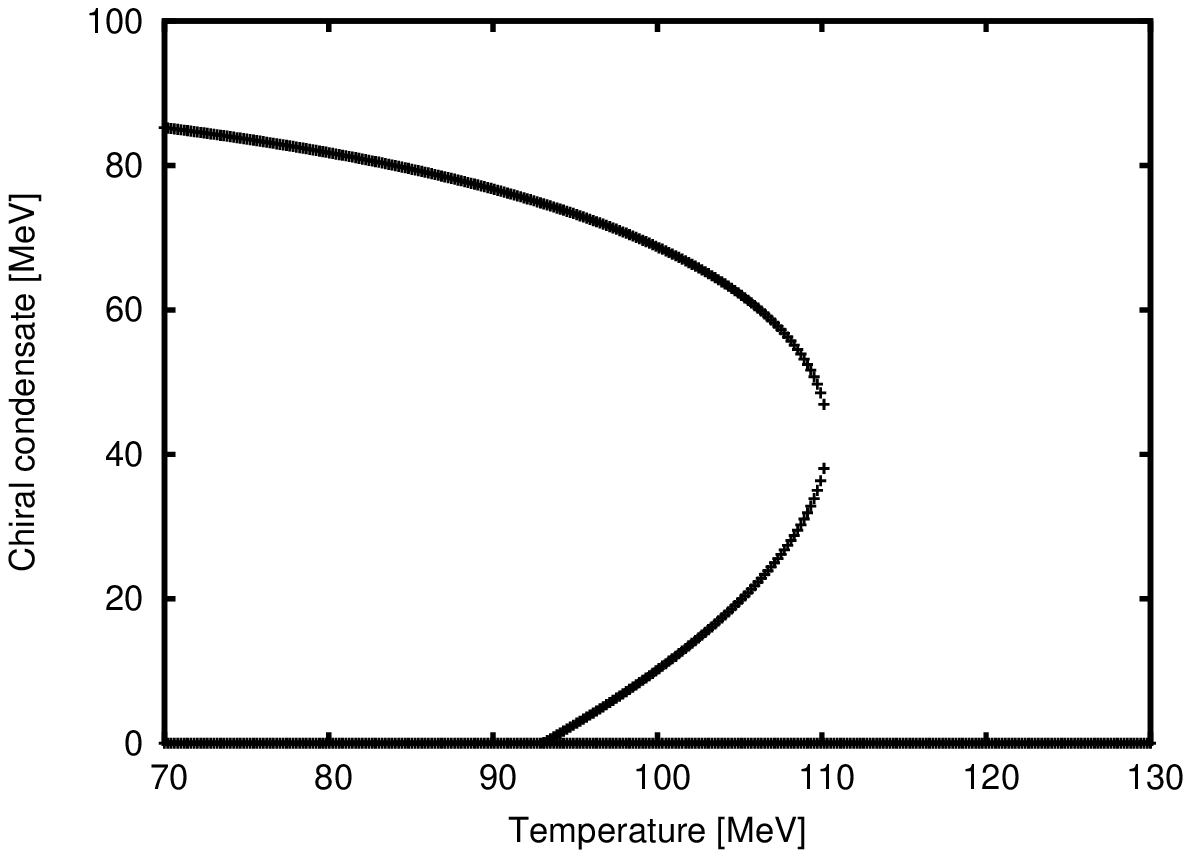}
        \caption{
            At one loop, including the thermal contributions from
            both fermions and bosons, but not vacuum contributions:
            (left panel) the phase diagram in the $\mu-T$ plane, where the
            solid line represents a first-order phase transition,
            and
            (right panel) the solutions of the gap equation plotted as a
            function of $T$ at $\mu = 0$.
            \label{fig:1-loop_chiral-limit_no-vacuum}
        }
    \end{figure}

    Next, let us omit the fermionic vacuum fluctuations again, and instead
    study the effects of including the thermal fluctuations of the mesons:
    \beq
        \mathcal F =
            \mathcal F_b^\mathrm{tree}
          + \mathcal F_b^T
          + \mathcal F_f^T\;.
    \eeq
    This situation was investigated by Bowman and Kapusta in
    Ref.\ \cite{Bowman:2008kc}.  In the chiral limit, they found that
    there is a first-order phase transition extending all the way from
    the $\mu$ axis to the temperature axis, with a critical temperature
    of $T_c \sim 140$ MeV at $\mu=0$.  We show this in
    Fig.\ \ref{fig:1-loop_chiral-limit_no-vacuum}, where we also see
    that we get a lower critical temperature of about
    110 MeV. The critical chemical
    potential is for $ T=0$ is $\mu_c=357$ MeV.


\begin{figure}[htb]
        \includegraphics[width=7cm]{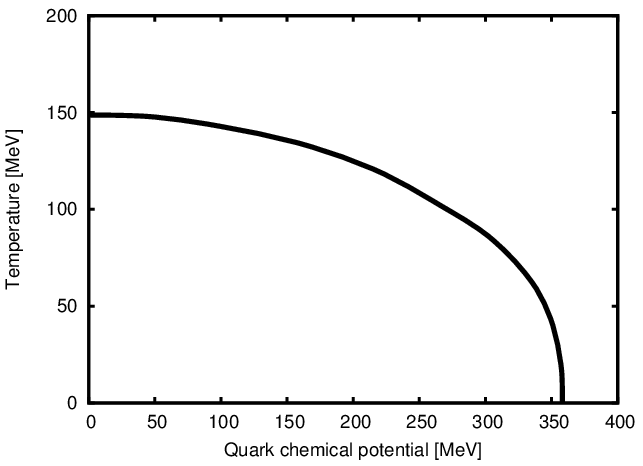}
        \includegraphics[width=7cm]{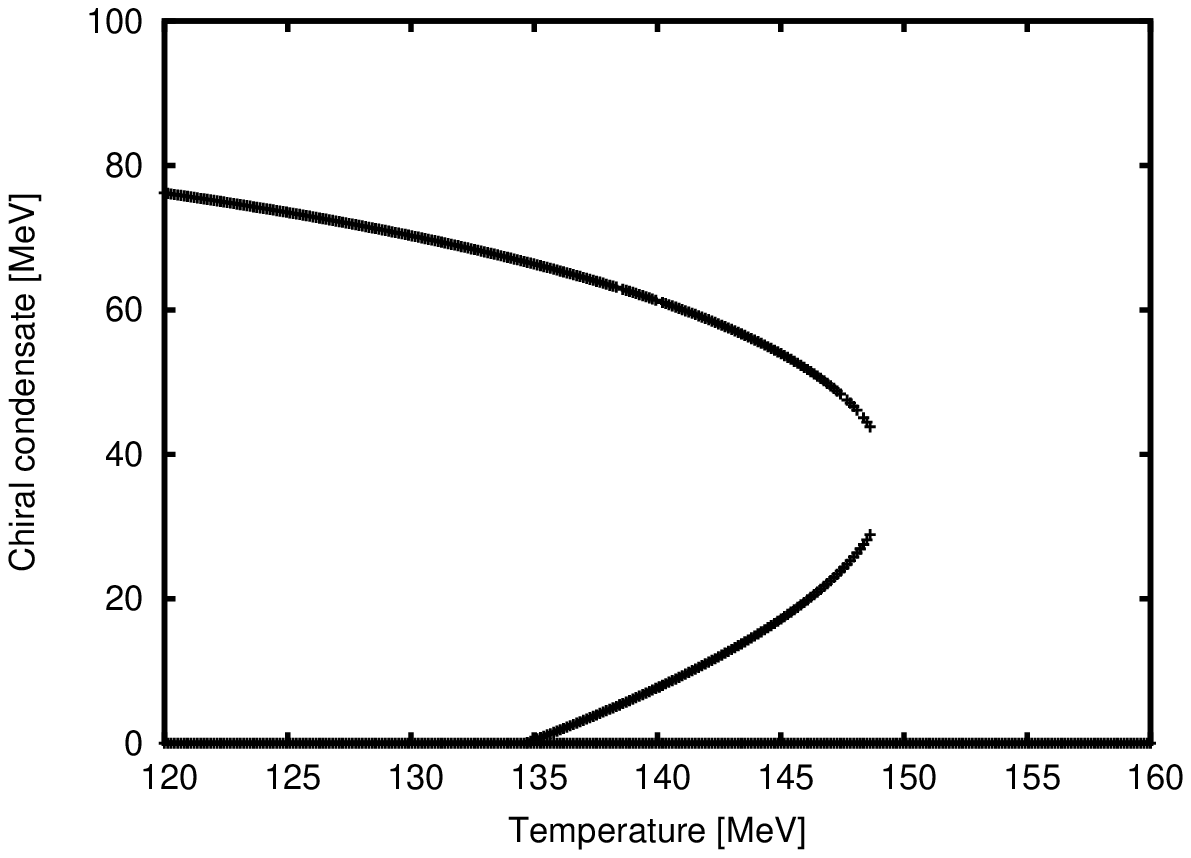}
        \caption{
            At one loop, including all contributions from
            both fermions and bosons:
            \emph{(left panel)} the phase diagram in the $\mu-T$ plane, where 
            the
            solid line represents a first-order phase transition,
            and
            \emph{(right panel)} the solutions of the gap equation plotted as a
            function of $T$ at $\mu= 0$.
            \label{fig:1-loop_chiral-limit_full}
        }
    \end{figure}

    We have now seen that by including $\mathcal F_f^\mathrm{vac}$ 
    the first-order phase transition turns into the second-order 
    transition one would expect at high temperatures, and that by
    taking $\mathcal F_b^T$ into account we pushed the critical
    temperature down to the vicinity of $T_c^\mathrm{QCD}$.
    It is therefore tempting to guess that by including all contributions,
    \beq
        \mathcal F =
            \mathcal F_b^\mathrm{tree}
          + \mathcal F_b^\mathrm{vac}
          + \mathcal F_b^T
          + \mathcal F_f^\mathrm{vac}
          + \mathcal F_f^T,
    \eeq
    we will get a second-order phase transition near $T = 150$ MeV.
    Unfortunately, this turns out not to be the case.
    As Fig.\ \ref{fig:1-loop_chiral-limit_full} shows, the critical
    temperature is 148 MeV, 
    but the phase transition is still first
    order---in fact it is even more strongly so.  Part of the reason
    for this is the fact that the vacuum energy of the mesons
    has opposite sign from the vacuum energy of the quarks and so
    the contributions cancel to some extent .The critical chemical
    for potential for $T=0$ is $\mu_c=357$ MeV.

\subsection{Physical point}
    In the chiral limit, $v$ is exactly zero in the chirally restored
    phase.  If we let $h > 0$, chiral symmetry is explicitly broken
    in the Lagrangian.  Then, the chiral condensate never vanishes
    completely, but approaches zero asymptotically at high temperature
    and/or large baryon chemical potential.
    Furthermore, the pions acquire mass and become pseudo Nambu--Goldstone
    bosons. 

    At the physical point, we use the vacuum values
    \beqa
        m_\sigma &=& 600\textrm{ MeV}\;,\\
        m_\pi    &=& 139\textrm{ MeV}\;,\\
        m_q      &=& 300\textrm{ MeV}\;, \\
        v        &=& f_\pi = 93\textrm{ MeV}\;,
    \eeqa
    and using Eqs.\ (\ref{eq:vacuum_m0})--(\ref{eq:vacuum_Lambda}) 
    we find the parameters
    \beqa
        m_0^2+\chi &=& -(388.611\textrm{ MeV})^2\;, \\
        \lambda &=& 118.17\;, \\
        g &=& 3.2258\;,\\
        h &=& (121.573\textrm{ MeV})^3\;,\\
        \Lambda &=& 363.92\textrm{ MeV}\;.
    \eeqa
        
    Let us look at what happens to the phase transition in the case
    when only thermal contrubtions to the free energy are taken
    into account.
    
    \begin{figure}[htb]
        \includegraphics[width=7cm]{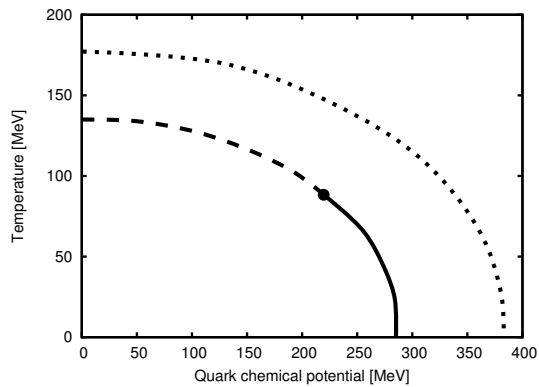}
       
        \caption{
            At one loop, including all contributions from
            both fermions and bosons (upper curve) and thermal contributions
only (lower curve).
            \label{fig:1-loop_physical-point_full}
        }
    \end{figure}

    This is shown in Fig~\ref{fig:1-loop_physical-point_full} (lower curve).
    The line of first-order transition that starts at 
    $\mu_c=287\,{\rm MeV}$ and $T=0$
    ends at a critical point at $\mu=220\,{\rm MeV}$ and $T=90\,{\rm MeV}$.
    In comparison, 
    Scavenius \cite{Scavenius:2000qd} {\it et al.} obtain
    $\mu=207\,{\rm MeV}$ and $T=99\,{\rm MeV}$
    for the position of the critical endpoint,
    while Bowman and Kapusta \cite{Bowman:2008kc} obtain
    the values
    $\mu=283\,{\rm MeV}$ and $T=75\,{\rm MeV}$. 

    If we include the bosonic as well as the fermionic vacuum contributions,
    i.e. we include all contributions in the model, the critical endpoint
    disappears. The transition is a crossover in the whole $\mu-T$ plane.
    This is shown in Fig~\ref{fig:1-loop_physical-point_full} as well for 
    comparison (upper curve). Only for lower values of $m_{\pi}$ will the
    critical endpoint appear.

\section{Summary and outlook} \label{sec:conclusion}
In the present paper, we have applied optimized perturbation theory
to the quark-meson model to study the chiral transition in the $\mu-T$
plane. The properties of the phase transition - the order and the critical
temperature - depend crucially on the the approximations made.
The one-loop calculation in the chiral limit yields
either a first-order or a second-order transition in the whole
$\mu-T$ plane depending on whether one includes the fermion
vacuum fluctuation term or not.
This is in contrast to most NJL-type model calculations~\cite{buballa} and
functional
renormalization group calculations
that predict a curve of first-order transitions
starting at $T=0$ ending at at critical 
point~\cite{bj,Schaefer:2006ds,Herbst:2010rf}.
Taking a model seriously means including the effects of all its degrees
of freedom. It therefore questionable to throw out some terms unless
one can show that they are not important. 
For example, the effective potential is 
receiving contributions from vacuum fluctuations from 
all energies scales up to the ultraviolet 
cutoff $\Lambda$ of the effective theory. This is the region of
validity of the low-energy effective theory
and it is therefore not obvious that they can be omitted.
In view of this, it is perhaps somewhat surprising that the critical endpoint
disappears all together at the physical point and we are left with a 
crossover in the whole $\mu-T$ plane.

We close with a few remarks concerning future work.
Optimized perturbation theory was used at the two-loop level
in $\phi^4$-theory to study the phase transition and it was shown that
it correctly reproduces the second-order nature at zero baryon chemical
potential~\cite{Chiku:2000eu}. It would therefore be of interest to
extend the two-loop calculations to the quark-meson model at finite $\mu$
along the lines of Ref.~\cite{Fraga:2009pi}.
Secondly, one could apply OPT to the case $N_f=3$, where we know the
transition in the chiral limit is first order.
Dialing the strange quark mass from $m_s=\infty$ to $m_s=0$  
interpolate between the chiral limit for $N_f=2$ and $N_f=3$.
Somewhere along the way, the transition then changes from second order
to first order and it would be interesting to find the critical value
for the strange quark mass, $m_s^*$ where this happens.

\appendix

\section{One-loop sum-integrals}  \label{sec:sumint}
    We need the following bosonic sum-integrals
    \beqa
        \sumint_P \log\left(P^2 + m^2\right) &=&
            {1\over(4\pi)^2} \left(\frac{\Lambda}{m}\right)^{2\epsilon}
            \left[
                - \frac{e^{\gamma \epsilon} \Gamma(1+\epsilon)}
                       {\epsilon (1-\epsilon) (2-\epsilon)} m^4
                - J_0(\beta m) T^4
            \right]\;,
        \\
        \sumint_P {1\over P^2 + m^2} &=&
            {1\over(4\pi)^2} \left(\frac{\Lambda}{m}\right)^{2\epsilon}
            \left[
                - \frac{e^{\gamma \epsilon} \Gamma(1+\epsilon)}
                       {\epsilon (1-\epsilon)} m^2
                + J_1(\beta m) T^2
            \right]\;.
    \eeqa
    The temperature-dependent 
term is expressed as integrals over the Bose-Einstein
    distribution function:
    \beqa
        J_n(x) &=&
            \frac{4 e^{\gamma\epsilon} \Gamma(\frac{1}{2})}
                 {\Gamma(\frac{5}{2} - n - \varepsilon)}
            x^{4-2n}  \int_0^\infty d t \;
            \frac{t^{4-2n-2\epsilon}}{\sqrt{t^2 + 1}}
            {1\over e^{x\sqrt{t^2 + 1}} - 1}\;.
    \eeqa
    These integrals satisfy the recursion relation 
    \beq
        \frac{d J_n}{d x} =
            2 \epsilon \frac{J_n(x)}{x}
            - 2 x J_{n+1}(x)\;.
    \eeq

    We need the following fermionic sum-integrals:
    \beqa\nonumber
        \sumint_{\{P\}}
            {\rm tr} \log(i \egamma_\mu  P_\mu + m - \mu \egamma_4)
        &=&
            {1\over(4\pi)^2} \left(\frac{\Lambda}{m}\right)^{2\epsilon}
            \bigg[
                - 2 \frac{e^{\gamma \epsilon} \Gamma(1+\epsilon)}
                         {\epsilon (1-\epsilon) (2-\epsilon)} m^4
                + [K_0^+(\beta m, \beta \mu) + K_0^-(\beta m, \beta \mu)] T^4
            \bigg]\;.
\\ &&
    \eeqa
    Here, the temperature-dependent term is 
expressed as integrals over the Fermi-Dirac
    distribution function:
    \beqa
        K_n^\pm(x, y)& =&
            \frac{4 e^{\gamma\epsilon} \Gamma(\frac{1}{2})}
                 {\Gamma(\frac{5}{2} - n - \varepsilon)}
            x^{4-2n}  
\int_0^\infty d t \;
            \frac{t^{4-2n-2\epsilon}}{\sqrt{t^2 + 1}}
            {1\over e^{x\sqrt{t^2 + 1} \pm y} + 1}.
    \eeqa
    The functions $K^{\pm}_n(x,y)$
satisfy the same recursion relation as $J_n(x)$,


\begin{thebibliography}{}
%
%
\bibitem{pisarski}
R.~D. Pisarski and F. Wilczek, 
Phys. Rev. D {\bf 29} (1984) 338.
\bibitem{stephanov}
M.~A. Stephanov, PoS LAT2006, (2006) 024.


\bibitem{qcdmu}
S. D. H. Hsu and M. Schwetz, Phys. Lett. B {\bf 432}, 203 (1998).

\bibitem{buballa}M. Buballa, Phys. Rept. {\bf 407}, (2005), 205. 

\bibitem{hotqcd}
  A.~Bazavov {\it et al.},
  Phys.\ Rev.\  D {\bf  80}, (2009), 014504.

\bibitem{wuppertal}S. Borsanyi {\it et al.}, 
JHEP 1011, (2010), 077 
\bibitem{Philipsen:2010gj}
O. Philipsen.
\newblock {arXiv:1009.4089} (2010). 

\bibitem{Chiku:1998kd}
S.~Chiku and T.~Hatsuda.
\newblock {Phys. Rev.} D58 (1998), 076001.


\bibitem{lenaghan}
J. T. Lenaghan and  D. H. Rischke, J. Phys. G {\bf 26}, (2000), 431;
J. T. Lenaghan, D. H. Rischke, and J. Schaffner-Bielich,
Phys. Rev. D {\bf 62}, (2000), 085008.


\bibitem{Chiku:2000eu}
S. Chiku.
\newblock {Prog. Theor. Phys.} {\bf 104}, (2000), 1129.

\bibitem{Herpay:2006vc}
T.~Herpay and Zs. Szep.
\newblock { Phys. Rev.} D {\bf 74}, (2006), 025008.

\bibitem{Farias:2008fs}
R.~L.~S. Farias, G.~Krein, and R.~O. Ramos.
\newblock { Phys. Rev.} D {\bf 78}, (2008), 065046.

\bibitem{Scavenius:2000qd}
O.~Scavenius, A.~Mocsy, I.~N. Mishustin, and D.~H. Rischke,
\newblock { Phys. Rev.} C {\bf 64} (2001), 045202.

\bibitem{kneur}
J.-L Kneur, M. B. Pinto, and R. O. Ramos, 
Phys. Rev. C {\bf 81}, (2010), 065205.

\bibitem{costa2}
P. Costa, M.C. Ruivo, and C.A. de Sousa, 
Phys. Rev. D {\bf 77}, (2008), 096001.


\bibitem{kahara}
T. Kahara and K. Tuominen,
Phys. Rev. D {\bf 78}, (2008), 034015; ibid D {\bf 82}, (2010), 114026.
\bibitem{nickel}
D. Nickel, Phys.Rev. D {\bf 80}, (2009), 074025.

\bibitem{costa}	
P. Costa, M. C. Ruivo, C. A. de Sousa, and H. Hansen,
Symmetry {\bf 2}, (2010), 1338.

\bibitem{jakobi}A. Jakovac, A. Patkos, Z. Szep, and P. Szepfalusy, 
Phys. Lett. B {\bf 582}, (2004), 179.


\bibitem{mocsy}A. Mocsy, I. N. Mishustin, and P. J. Ellis, 
Phys. Rev. C {\bf 70}, (2004), 015204.

\bibitem{bj}
B.-J. Schaefer and J. Wambach, Nucl. Phys.  A {\bf 757}, (2005), 479. 

\bibitem{Schaefer:2006ds}
B.-J. Schaefer and J. Wambach,
\newblock { Phys. Rev.} D {\bf 75}, (2007), 085015.



\bibitem{Kovacs:2006ym}
P.~Kovacs and Zs. Szep.
\newblock { Phys. Rev.} D {\bf 75}, (2007), 025015.



\bibitem{jocke}B.-J. Schaefer and  M. Wagner, 
Phys. Rev. D {\bf 79}, (2009), 014018. 

\bibitem{Bowman:2008kc}
E.~S. Bowman and J.~I. Kapusta,
\newblock { Phys. Rev.} C {\bf 79}, (2008), 015202.

\bibitem{Skokov:2010sf}
V.~Skokov, B.~Friman, E.~Nakano, K.~Redlich, and B.-J. Schaefer,
\newblock {Phys. Rev.} D {\bf 82} (2010), 034029.

\bibitem{Jakovac:2010uy}
A.~Jakovac and Zs. Szep, Phys. Rev. D {\bf 82}, (2010), 125038. 


\bibitem{koch}
L. Ferroni, V. Koch, and M. B. Pinto, Phys. Rev. C {\bf 82}, (2010) 055205.
\bibitem{Fraga:2009pi}
E.~S. Fraga, L.~F. Palhares, and M.~B. Pinto.
\newblock { Phys. Rev.} D79 ,(2009), 065026.

\bibitem{fuku}
K. Fukushima, 
Phys. Lett. B {\bf 591}, (2004), 277.


	

\bibitem{Schaefer:2007pw}
B.-J. Schaefer, J.~M. Pawlowski, and J. Wambach.
\newblock { Phys. Rev.} D {\bf 76}, (2007), 074023.
\bibitem{mats}
K. Kashiwa, H. Kouno, and M. Matsuzaki, and M. Yahiro
Phys. Lett. B {\bf 662}, (2008), 26.

\bibitem{mao}H. Mao, J. Jin, and M. Huang,
J. Phys. G {\bf 37} (2010), 035001. 

\bibitem{gupti}
U. S. Gupta and V. K.Tiwari, Phys. Rev. D {\bf 81}, (2010), 054019.

\bibitem{Marko:2010cd}
  G.~Marko and Zs. Szep, Phys. Rev. D {\bf 82}, (2010), 065021.



\bibitem{Herbst:2010rf}
T.~K. Herbst, J.~M. Pawlowski, and B.-J. Schaefer.
Phys. Lett. B {\bf 696}, 58 (2011). 

\bibitem{nonlocal}
A. E. Radzhabov, D. Blaschke, M. Buballa, and M. K. Volkov, 
e-Print: arXiv:1012.0664 [hep-ph]. 

\bibitem{alford}
M. G. Alford, A. Schmitt, and K. Rajagopal,
Rev. Mod. Phys. {\bf 80}, (2008), 1455.
\bibitem{mizher}
A.~J. Mizher and E.~S. Fraga.
\newblock
\newblock Nucl. Phys. A 831, (2009) 91.



\bibitem{yuk}
  V.~I.~Yukalov,
  Teor.\ Mat.\ Fiz.\ {\bf 26} (1976), 403.
\bibitem{steve}
  P.~M.~Stevenson,
  Phys.\ Rev. D {\bf 23} (1981), 2916.
\bibitem{duncan1}
  A.~Duncan and M.~Moshe,
  Phys.\ Lett.\ {\bf B 215} (1988), 352.
\bibitem{duncan2}
  A.~Duncan and H.~F.~Jones,
  Phys.\ Rev.  D {\bf 47} (1993), 2560.
\bibitem{kleinert}
  H.~Kleinert, {\it Path Integrals in Quantum Mechanics, Statistics, and Polymer Physics},
  2nd edition, World Scientific Publishing Co., Singapore, (1995).
\bibitem{russ}
  A.~N.~Sisakian, I.~L.~Solovtsov and O.~Shevchenko,
  Int.\ J.\ Mod.\ Phys.\ {\bf A 9} (1994), 1929.
\bibitem{janke}
  W.~Janke and H.~Kleinert,
  Phys.\ Rev.\ Lett.\ {\bf 75}, (1995), 2787.


\bibitem{karsch}
  F.~Karsch, A.~Patkos and P.~Petreczky,
  Phys.\ Lett.\ {\bf B 401}, (1997), 69. 





\newblock { Phys. Lett.} B401, (1997), 69.

\bibitem{Andersen:2000yj}
J.~O. Andersen, E. Braaten, and M. Strickland,
\newblock { Phys. Rev.} D {\bf 63}, (2001), 105008.





















\end{thebibliography}
\end{document}